\documentclass[twocolumn,aps,prl,showpacs,amsmath,superscriptaddress,longbibliography,notitlepage]{revtex4-1}
\usepackage{amssymb}
\usepackage{mathrsfs}
\usepackage{graphicx}
\usepackage{float}
\usepackage[caption=false]{subfig}
\usepackage{epstopdf}
\usepackage[normalem]{ulem}
\usepackage{color}
\usepackage{bm}
\usepackage{verbatim}

\begin{document}

\def\Mu{\textcolor{red}}
\def\CH{\textcolor{magenta}}
\def\CHH{\textcolor{cyan}}
\def\GJ{\textcolor{blue}}

\title{Emergent Fermi surface in a many-body non-Hermitian Fermionic chain}

\author{Sen Mu}
\affiliation{Department of Physics, National University of Singapore, Singapore 117542}
\author{Ching Hua Lee}
\email{calvin-lee@ihpc.a-star.edu.sg}
\affiliation{Department of Physics, National University of Singapore, Singapore 117542}
\affiliation{Institute of High Performance Computing, A*STAR, Singapore 138632}
\author{Linhu Li}
\affiliation{Department of Physics, National University of Singapore, Singapore 117542}
\author{Jiangbin Gong}
\email{phygj@nus.edu.sg}
\affiliation{Department of Physics, National University of Singapore, Singapore 117542}

\date{\today}

\begin{abstract}
 {Quantum degeneracy pressure (QDP) underscores the stability of matter and is arguably the most ubiquitous many-body effect}. {The associated Fermi surface (FS) has broad implications for physical phenomena, ranging from electromagnetic responses to entanglement entropy (EE) area law violations. 
Given recent fruitful studies in condensed-matter physics under effectively non-Hermitian descriptions, it becomes urgent to study how QDP and many-body interactions interplay with non-Hermitian effects. Through a prototypical critical 1D fermionic lattice with asymmetric gain/loss, a real space FS is shown to naturally emerge, in addition to any existing momentum space FS. We also carefully characterize such real space FS with the EE, by a renormalized temperature that encapsulates the interplay of thermal excitations and non-hermiticity.}
Nearest neighbor repulsion is also found to induce competing charge density wave (CDW) that may erode the real space FS.  The underlying physics surrounding criticality and localization is further analyzed with complex flux spectral flows. Our findings can be experimentally demonstrated with ultracold fermions in a suitably designed optical lattice.

\end{abstract}

\maketitle

 {In a wide variety of systems such as quantum Hall liquids, superconductors and neutron stars}, it is the emergent many-body effects, rather than single-particle behavior, that give rise to their respective signature properties. {{Arguably quantum} degeneracy pressure (QDP) represents} the most ubiquitous many-body effect, {where} the Pauli exclusion principle 
underscores both the rigidity of everyday-life solids \cite{Dyson1967sta1,Dyson1967sta2,Dyson1967sta3} and the stability of neutron stars \cite{Oppenheimer1939}. A {primary} consequence of QDP is the formation of a Fermi surface (FS), which bounds a sea of impenetrable fermions in optimal energetic configuration.
Dictating the available quasi-particle excitations and semiclassical contours, the FS shape crucially controls transport, magnetization and optical properties~\cite{zhao2015anisotropic,zhu2015quantum,lee2015negative,morimoto2016semiclassical,grushin2016inhomogeneous,moll2016transport,de2017quantized,hamamoto2017nonlinear,lee2019enhanced}.
As an extended critical region, a FS also violates the celebrated area law for entanglement entropy (EE), {whose
deep relation with many-body couplings have spurred the study of holographic duality} \cite{Wilczek1994,Cardy2004,Wolf2006,Ryu2006holo,Gioev2006,Senthil2008,Lee2008sfs,Eisert2010area,Swingle2010ent,qi2013exact,gu2016holographic,you2016entanglement}. 

{Non-Hermitian descriptions of condensed-matter systems \cite{Duan2017,JapanPRL,FuPRB,FuPRL2,hamazaki2019non,nakagawa2018non,zhao2019non,HuPRB,WangPRL2019,Wu2019science,Li2019geometric,nakagawa2018non,zhao2019non,el2018non,gong2018topological,zhong2018winding,kawabata2019topological,kawabata2019classification,yoshida2019non,liu2019topo,Lee2016nonH,Martinez2018nonH,Yao2018edge,Yao2018nonH2D,Kunst2018biorthogonal,Lee2019anatomy,Lee2019hybrid,Lee2018tidal,Yokomizo2019nonbloch,Longhi2019prr,zhang2019floskin,longhi2019nhquasi,jiang2019interplay} have provided an effective and fruitful framework to account for inelastic collisions \cite{JapanPRL}, disorder effects \cite{FuPRB,FuPRL2,hamazaki2019non}, and system-environment couplings \cite{nakagawa2018non,zhao2019non,HuPRB,WangPRL2019}.  This research avenue has extended the domain of condensed-matter physics with inspiring insights. It is therefore necessary and urgent to study the implications of non-Hermiticity for QDP.  In particular, nonreciprocal hopping in a lattice system defines a preferred pumping direction, thereby causing all eigenstates at the single-particle level to accumulate at the boundaries. However, this non-Hermitian phenomenon, coined the {non-Hermitian} skin effect~\cite{Lee2016nonH,Martinez2018nonH,Yao2018edge,Yao2018nonH2D,Kunst2018biorthogonal,Lee2019anatomy,Lee2019hybrid,Lee2018tidal,Yokomizo2019nonbloch,Longhi2019prr,zhang2019floskin,longhi2019nhquasi,jiang2019interplay},  
cannot possibly persist in the presence of QDP and many-body interactions, which will at the very least prohibit multiple occupancy at the boundaries.}

In this work, we show how the non-Hermitian skin effect can naturally yield a Fermi sea in real space. Because spatial particle accumulation is not physically identical with the Fermi sea condensation in energy space, we next justify how this emergent accumulation corresponds to a bona-fide FS at the level of the EE, characterized by a renormalized temperature depending on both physical temperature and hopping asymmetry.  {Furthermore, we observe the erosion of the emergent FS by the charge-density-wave (CDW) arising from the nearest neighbour (NN) repulsion}. Finally, we visualize these interplay in terms of spectral flows, and suggest a cold-atom setup for future experimental demonstration.
\vspace{0.15cm}

\begin{figure}
\includegraphics[width=1.0\linewidth]{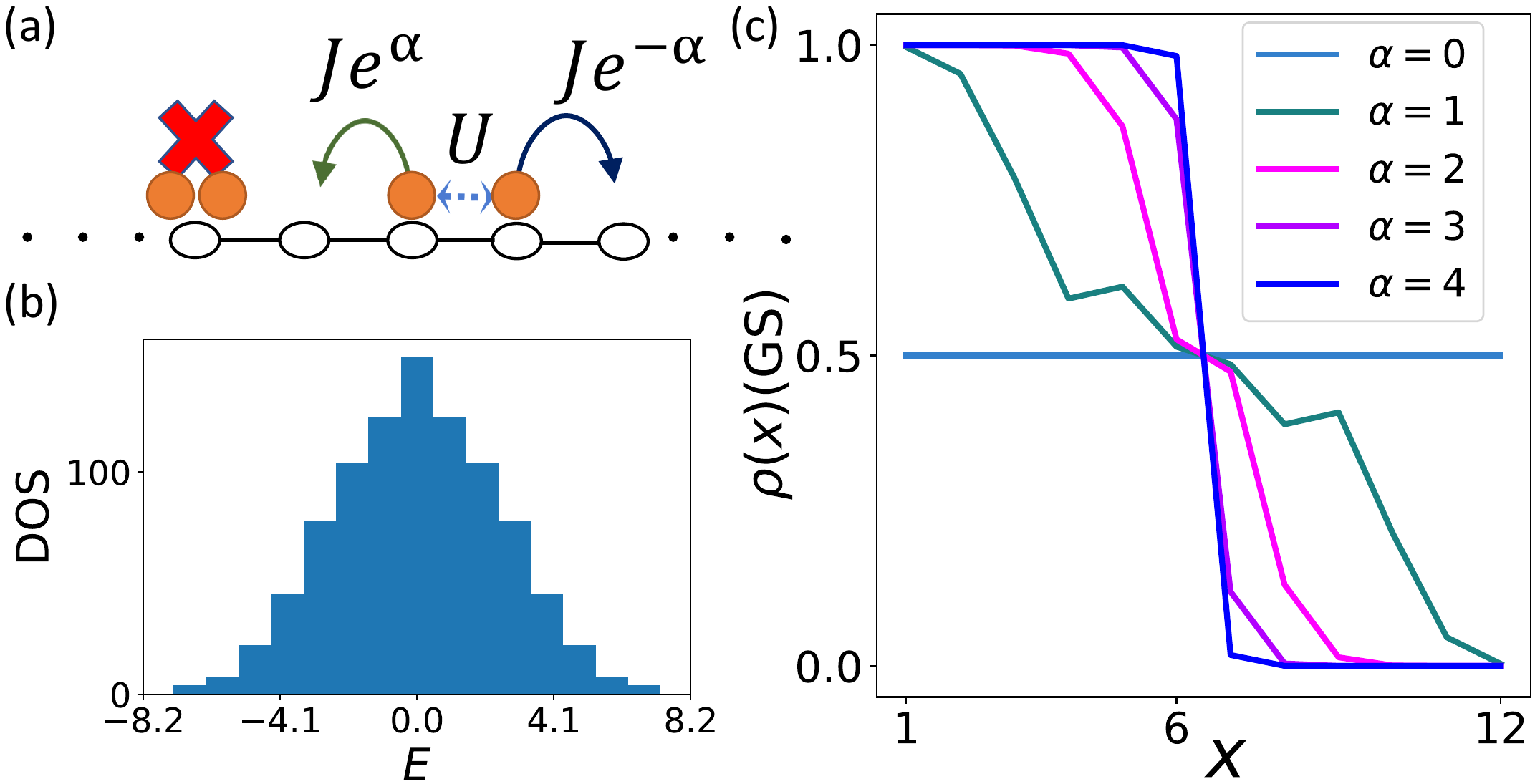}
\caption{(a) Our fermionic chain Eq.~(\ref{pm}) with asymmetric gain/loss $Je^{\pm \alpha}$ and NN interactions $U$. (b) Its OBC density of states (DOS) at $U=0$, corresponding to a real gapless spectrum. (c) Spatial density of the GS at $U=0$ and half filling $L=2n=12$, which 
reveals a real space FS that becomes sharper with increasing $\alpha$ asymmetry.}
\label{fig_model}
\end{figure}

\noindent{\it Interacting fermions with asymmetric gain/loss.--} We consider a minimal model that captures the interplay between asymmetric non-Hermitian gain/loss and two types of many-body effects: (i) fermionic QDP and (ii) NN repulsion. As illustrated in Fig.~\ref{fig_model}(a), it consists of spin-polarized repulsive fermions hopping along a chain of length $L$ with open boundaries conditions (OBCs):
\begin{equation}
  H = \sum_{x=1}^{L-1}\{J(e^{\alpha}c^{\dagger}_{x}c_{x+1}+e^{-\alpha}c^{\dagger}_{x+1}c_x)+Un_xn_{x+1}\},
  \label{pm}
\end{equation}
where  $c^{\dagger}_x$/$c_x$ is the fermion creation/annihilation operator at site $x$ and $n_x=c^{\dagger}_xc_x$ is the corresponding fermion number operator. Two fermions are forbidden from occupying the same site, and will incur an energy penalty of $U>0$ if they occupy adjacent sites. {The nonreciprocal left/right hoppings $Je^{\pm \alpha}$ can be understood as asymmetric gain/loss and are within reach of experiments} \cite{ghatak2019observation,helbig2019observation,xiao2019observation,hofmann2019reciprocal}. Unless otherwise stated, we shall assume half filling (presence of $n=L/2$ fermions).

The OBC spectrum of this simple ansatz system is always real and gapless (Fig.~\ref{fig_model}{(b)}), as seen through the spatially inhomogeneous similarity transform $c^\dagger_x\rightarrow c^\dagger_x e^{x\alpha}$, which eliminates the $e^{\pm \alpha}$ factors in Eq.~(\ref{pm}) and keeps $n_i$ invariant \cite{Zhang2013}. As such, familiar concepts like the ground state (GS) and energy gaps remain applicable.
Since the spectrum is agnostic to the non-Hermitian asymmetry $\alpha$, the interplay between many-body effects and non-Hermiticity is only manifested at the \emph{eigenstate} level. Henceforth we focus on the eigenstates, except when studying broader implications on bulk-boundary spectral correspondences.
\vspace{0.15cm}

\noindent{\it Emergent real space Fermi surface.--} {Demarcating the occupied Hilbert space boundary, a} {FS plays a dual role to the entanglement cut and allows the EE to be expressed in a position-momentum symmetric manner \cite{its2005entanglement,its2009fisher,Lee2014PRPRP}. As such, a real space FS can co-exist on equal footing with an ordinary momentum space FS.}

{Here we show how a real space FS emerges naturally} from the interplay of non-Hermiticity and QDP. Consider first the case without NN repulsion ($U=0$). At the single-particle level, $H|_{U=0}$ possesses non-Bloch eigenstates $\psi_j(x)=\mathcal{N}_je^{-x\alpha}\sin\frac{\pi j x}{L+1}$ with corresponding eigenenergies $\varepsilon_j=2J\cos\frac{\pi j}{L+1}$ and normalization constants $\mathcal{N}_j$~\cite{suppmat}. For $\alpha>0$, the eigenstates are all exponentially localized at the left boundary $(x=1)$ with localization length $\alpha^{-1}$, as implied by the above-mentioned similarity transformation. Physically, they also represent the steady state solutions of a biased random walk on a bounded 1D chain.

Multiple fermions, however, will not be allowed to all localize at $x=1$ due to QDP. We characterize the spatial density of an $n$-fermion state $\psi_\mu(x_1,...,x_n)$ by $\rho_\mu(x)=\int |\psi(x_1,...,x_{n-1},x)|^2\,\prod_{i=1}^{n-1}dx_i$, and define the thermal-weighted density $\rho(x)={N \choose n}^{-1}\sum_\mu e^{-\beta E_\mu}\rho_\mu(x)$ with temperature $\beta^{-1}$, where $E_\mu$ is the energy of $\psi_\mu$. For $\beta\rightarrow \infty$, we obtain the GS density $\rho(x)(\text{GS})$.

From Fig.~\ref{fig_model}(c),  $\rho(x)(\text{GS})$ is seen to be spatially uniform for reciprocal hopping ($\alpha=0$). 
As $\alpha$ increases, we observe a competition between asymmetric hoppings $e^{\pm \alpha}$ and QDP: While even a tiny $\alpha$ has the propensity to localize each fermion towards the left boundary, the 
 {QDP} forces all the $n$ fermions to forbid another from occupying the same site. As such, $\rho(x)(\text{GS})$ is symmetric about the $n$ and $(n+1)$-th site, with profile controlled by the exponential tail of a single-fermion eigenstate. In the extremely asymmetric limit of $|\alpha|\rightarrow \infty$, hoppings become unidirectional, and the density profile becomes a jump discontinuity. Note that this nontrivial compromise of QDP and asymmetric gain/loss only applies to fermions, since multiple bosons will be allowed to condense macroscopically at one boundary, just like isolated bosons.

From Fig.~\ref{fig_model}(c), $\rho(x)(\text{GS})$ very closely resembles to a Fermi-Dirac spatial profile of the form $\rho_{\rm FD}(x)=(1+e^{\Lambda(x-n-1/2)})^{-1}$, with $\Lambda\sim 4\alpha$ rigorously derivable from the Slater determinant~\cite{suppmat}. Attractive as this identification looks, $\Lambda$ \emph{cannot} represent an effective inverse temperature because it is conjugate to position, not energy. To transcend this subtlety and determine the exact sense in which we have an emergent real space FS, a universal recourse is the entanglement entropy, whose scaling behavior reveals both the temperature and FS properties. Conformally transforming standard results \cite{Cardy2004,Cardy2009}, the EE $S_{\rm ent,\beta}$ of a finite critical half-filled 1D system scales like
\begin{eqnarray}
S_{\rm ent,\beta} \sim \frac{c}{6}\log\left[\frac{\beta}{\pi}\sinh\frac{\pi L_A}{\beta}\right]+\mathrm{const.},
\label{obc_ee_t}
\end{eqnarray}
where $\beta$ is the inverse temperature, $L_A$ is the position of the entanglement cut and $c=1$ is the central charge for our fermionic model. Its logarithmic form violates the area law of EE scaling, which states that the EE of a gapped 1D system should plateau beyond sufficiently large system size. In the following, we shall also investigate how this violation is further modified by non-Hermiticity $\alpha$.

To obtain the EE, note that non-Hermitian Hamiltonians, even those with real spectra, generically possess different left and right eigenvectors 
defined by $H^\dagger|\psi^L\rangle = \varepsilon^* |\psi^L\rangle $ and $H|\psi^R\rangle = \varepsilon |\psi^R\rangle $. As such, the density matrix can either be defined biorthogonally i.e. $[\rho^{RL}]_{\mu\nu}=|\psi^R_\mu\rangle\langle \psi^L_\nu|$, or with respect to only left or right eigenvectors i.e. $[\rho^{RR}]_{\mu\nu}=|\psi^R_\mu\rangle\langle \psi^R_\nu|$. In this work, we shall take the latter option because we are interested in the eigenstates themselves, rather than computing probability-conserving expectation values of observables \cite{herviou2019entanglement,ryu2019entanglement}. From $\rho^{RR}$, we can trace out degrees of freedom other than $A$ and define a reduced density matrix $\rho_A^{RR}=\mathrm{Tr}_{A^{\rm c}}|\psi^R\rangle\langle\psi^R|=\sum_r \lambda_r^2 |\psi^R_{r,A}\rangle\langle \psi^R_{r,A}|$, where $|\psi^R\rangle=\sum_r\lambda_r |\psi^R_{r,A}\rangle\otimes |\psi^R_{r,A^{\rm c}}\rangle$ is the Schmidt decomposition of a representative $|\psi^R\rangle$ and $A^{\rm c}$ is the complement of $A$. The von Neumann EE with respect to the entanglement cut separating $A$ and $A^{\rm c}$ is given by
\begin{eqnarray}
S_{\rm ent} &=& -\mathrm{Tr}[\rho^{RR}_A\mathrm{log}\rho^{RR}_A]\notag\\
&=&-\sum_r [\lambda_r\log\lambda_r + (1-\lambda_r)\log(1-\lambda_r)],
\label{sent}
\end{eqnarray}
with $A$ and $RR$ made implicit from now on.

\begin{figure}[t]
\includegraphics[width=1.0\linewidth]{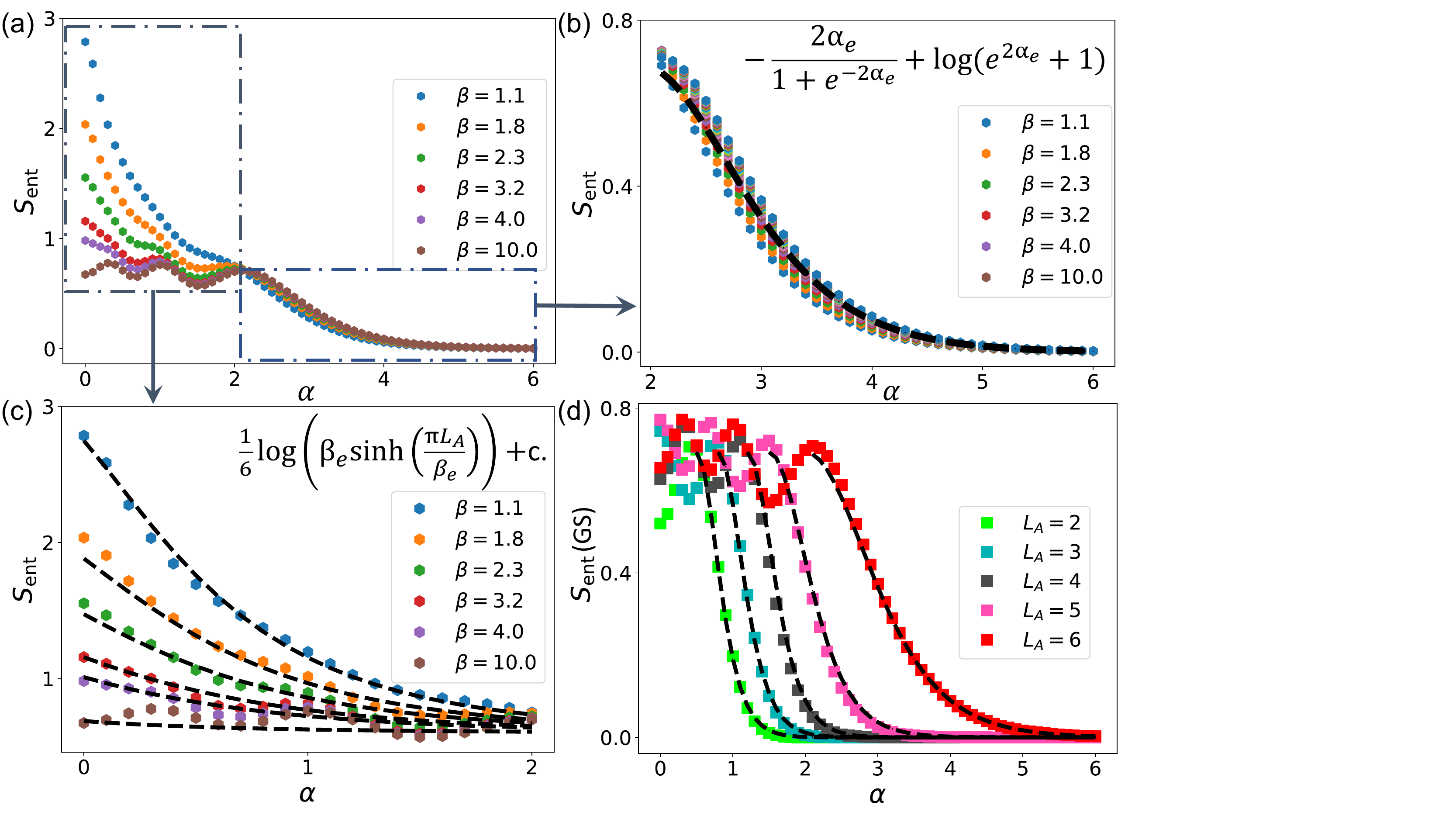}
\caption{EE behavior giving rise to renormalized temperature and hopping asymmetry. (a) OBC EE vs. $\alpha$ at different inverse physical temperatures $\beta$ with $L=2L_A=12$ at half filling. The $\beta$-independent $\alpha>2.1$ regime and logarithmically scaling $\alpha<2.1$ regime are detailed in (b,c) respectively. (b) Large $\alpha$ regime fitted to Eq.~(\ref{2q_ee}) with renormalized asymmetry $\alpha_{\rm e}=\eta(\alpha-\alpha_0)$, where $\alpha_0=2.1,\eta=1$. (c) Small $\alpha$ regime fitted to $S_{\rm ent,\beta_e}$ from Eq.~(\ref{obc_ee_t}) with renormalized temperature $\beta_e=\beta+0.75\alpha\beta+1.25\alpha^2$. (d) OBC EE vs. $\alpha$ at zero temperature for different subsystem lengths $L_A$ and $L=12$ at half filling. Eq.~(\ref{2q_ee}) still fits excellently, with $\alpha_0=\{0.5,0.8,1.2,1.5,2.1\}$ and $\eta=\{3.0,2.7,2.5,1.7,1.0\}$ for $L_A=\{2,3,4,5,6\}$.}
\label{fig_ee}
\end{figure}

From Fig~\ref{fig_ee}(a), $S_{\rm ent}$ behaves very differently across the regimes of weak and strong $\alpha$, demarcated at $\alpha_0\approx 2.1$ for the parameters ($L_A=L/2=12$) used. In the weak $\alpha$ regime, which is further elaborated in Fig~\ref{fig_ee}(c), $S_{\rm ent}$ varies strongly with both $\alpha$ and inverse temperature $\beta$, suggestive of their strong interplay. In the strong $\alpha$ regime (Fig~\ref{fig_ee}(b)), however, $S_{\rm ent}$  shows little dependence on $\beta$, suggesting that sufficiently strong hopping asymmetry $e^{\pm\alpha}$ generates a robust real space FS that dominates any smudging effect from the original thermal ensemble.
\vspace{0.15cm}

\noindent{\it Renormalized temperature and gain/loss asymmetry.--} Interestingly, the effect of the $\alpha$ in the weak asymmetry regime $\alpha<\alpha_0$ can be understood as a \emph{renormalization} of the effective temperature. Intuitively, asymmetric gain/loss pushes all fermions towards one side, decreasing their configurational freedom and hence increasing the cost of ''excitations''. At the level of EE, this reduction of freedom reduces the entanglement, mirroring the entanglement drop with decreased thermal excitations. This is substantiated by the fitted curves in Fig.~\ref{fig_ee}(c), where the EE ($L_A=L/2$) for $0<\alpha<\alpha_0$ is shown to agree very well~\footnote{The fit is excellent except at the oscillations at large $\beta$, which are finite size effects.} with its \emph{Hermitian} ($\alpha = 0$) expression Eq.~(\ref{obc_ee_t}), but at a renormalized inverse temperature
$\beta_{\rm e}=\beta+1.25\alpha^2+0.75\alpha\beta$.
The $\alpha^2$ and $\alpha\beta$ terms respectively represent the effective temperature suppression due to the real space FS and the mutual coupling between the real space  and momentum space Fermi seas. 
Substituting $\beta_{\rm e}$ into Eq.~(\ref{obc_ee_t}), we obtain an EE expression that violates the area law with an unconventional temperature dependence.

{In the strong asymmetry regime of $\alpha > \alpha_0$, the EE becomes almost temperature independent, indicating that the EE is dominated by the sharp real space FS. This can be further understood} through a 2-qubit model, where it is the asymmetry parameter $\alpha$ that becomes renormalized instead. Due to the sharp FS, we can approximate a generic state by  $|\psi\rangle\propto e^{\alpha_{\rm e}/2}|1_A0_{\bar A}\rangle+e^{-\alpha_{\rm e}/2}|0_A1_{\bar A}\rangle$, where $A,\bar A$ are the subsystems demarcated by an entanglement cut.
Taking a partial trace over the $\bar A$ qubit in the density matrix $\rho=|\psi\rangle\langle\psi|$, we obtain the reduced density matrix $\rho_A = e^{2\alpha_e}/(1+e^{2\alpha_e})|0_A\rangle\langle0_A|+1/(1+e^{2\alpha_e})|1_A\rangle\langle1_A|$, which possesses the EE
\begin{eqnarray}
S_{\rm ent,A}=-\frac{2\alpha_{\rm e}}{1+e^{-2\alpha_{\rm e}}}+\log(e^{2\alpha_{\rm e}}+1),
\label{2q_ee}
\end{eqnarray}
which from Fig.~\ref{fig_ee}(b) agrees very well with the actual EE of our model Eq.~(\ref{pm}) when $\alpha>\alpha_0$, with the renormalized $\alpha$ simply given by $\alpha_{\rm e}=\alpha-\alpha_0$. This validates our 2-qubit caricature in the $\alpha>\alpha_0$ regime, below which thermal excitations are sufficiently strong to break down this two-level picture and produce EE beyond $\log 2 \approx 0.69$.

At zero temperature, our 2-qubit model remains fully applicable even when the entanglement cut does not coincide with the real space FS, i.e. $L_A \neq n$, where $n=L/2$ in our case for half filling. Shown in Fig.~\ref{fig_ee}(d) are excellent fits of the EE $S_{\rm ent}$ with Eq.~(\ref{2q_ee}) for $L_A=2$ to $6$ ($n=6$), with details of $\alpha_{\rm e}$ given in the caption. For sufficiently large $\alpha$, $S_{\rm ent}$ drops sharply when the emergent real space FS is not aligned with the entanglement cut, implying that the FS harbors most of the entanglement. 

\begin{figure}[t]
\includegraphics[width=1.0\linewidth]{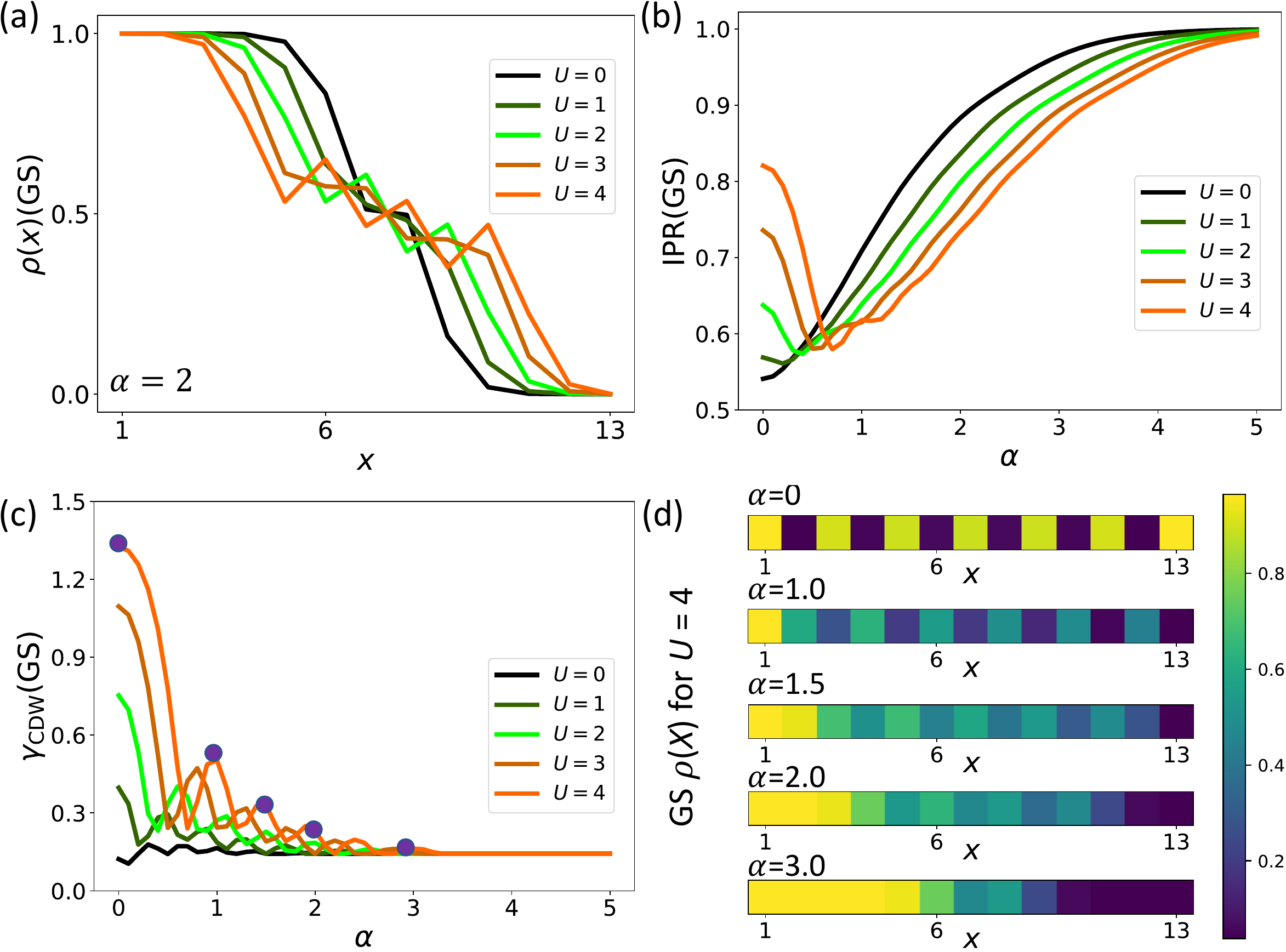}
\caption{NN interaction $U$ which induces CDW interplay with $\alpha$, all with $n=7$ and $L=13$. (a) Spatial GS density, with FS destabilized by increasing $U$. (b,c) $\rm IPR$/$\gamma_{\rm CDW}$ of the GS vs. $\alpha$, both of which are non-monotonic with $\alpha$. (d) GS spatial profiles at $U=4$ and various $\alpha$ as indicated by the purple dots in (c). As $\alpha$ increases, the CDW gives way to the FS, with the localization (IPR) minimized at an intermediate stage.}
\label{fig_u}
\end{figure}

\vspace{0.15cm}
\noindent{\it Competiting CDW and asymmetric gain/loss.--} We now turn on the NN repulsion $U$ and study how it can destabilize the emergent FS. Like QDP, nonzero $U>0$ also serves to counteract boundary mode accumulation through repulsion. Indeed as portrayed by Fig.~\ref{fig_u}(a), increasing $U$ smooths out the FS in a way naively reminiscent of decreasing the $\alpha$ asymmetry.

A closer examination of the individual eigenstate profiles reveals striking differences between QDP, which acts relentlessly on all fermions, and $U$ repulsion, which only assigns finite energy penalties. We consider the inverse participation ratio (IPR)~\footnote{This definition is identical to that in Hermitian cases, but alternative biorthogonal or SVD definitions exist \cite{herviou2019entanglement}.}, $\text{IPR}\,(\mu)=\frac1{n}\frac{\sum_x|\langle x|\psi_\mu \rangle|^4}{(\langle \psi_\mu |\psi_\mu \rangle)^2}$,
which reveals the real space locality of the $\mu$-th right eigenstate $\psi_\mu$: $\text{IPR}\,(\mu)=L^{-1}$ or $1$ in the extreme cases where $\psi_\mu$ is spatially uniform or localized on one site respectively. Focusing on the GS ($\mu = \rm{GS}$) which is minimally penalized by $U$, we observe an enigmatic trend in Fig.~\ref{fig_u}(b) where the IPR can vary non-monotonically with $\alpha$. In the $U=0$ limit, the IPR simply increases monotonically with $\alpha$ to its maximal value as the FS becomes sharper. But with nonzero NN repulsion $U$, the IPR actually dips before rising again, signifying a competing delocalizing influence. 

Intuitively, the NN repulsion can favor charge density waves (CDWs) because it repels adjacent fermions but allows them to accumulate freely as next-NNs. To check if this intuition corroborates with the non-monotonic IPR behavior, we compute the CDW imbalance parameter {$\gamma_{\rm CDW}(\mu)=\frac{2}{L}\sum_{x=1}^{L-1}|\rho_\mu(x+1)-\rho_\mu(x)|$} for $\mu=\rm{GS}$, which ranges from $0$ to $2$ depending on how closely the GS assumes a ferromagnetic or anti-ferromagnetic spatial density profile. From Fig.~\ref{fig_u}(c), it is evident that around the dip in IPR, the GS $\gamma_{\rm CDW}$ is indeed large even for moderately strong $U$. To completely interrogate this CDW behavior, we examine the density profiles of the $U=4$ case at various $\alpha$ (Fig.~\ref{fig_u}(d)). Indeed, the large IPR at small and large $\alpha$ are due to different reasons, namely CDW and FS localization respectively.
\vspace{0.15cm}

\begin{figure*}
\includegraphics[width=1.0\linewidth]{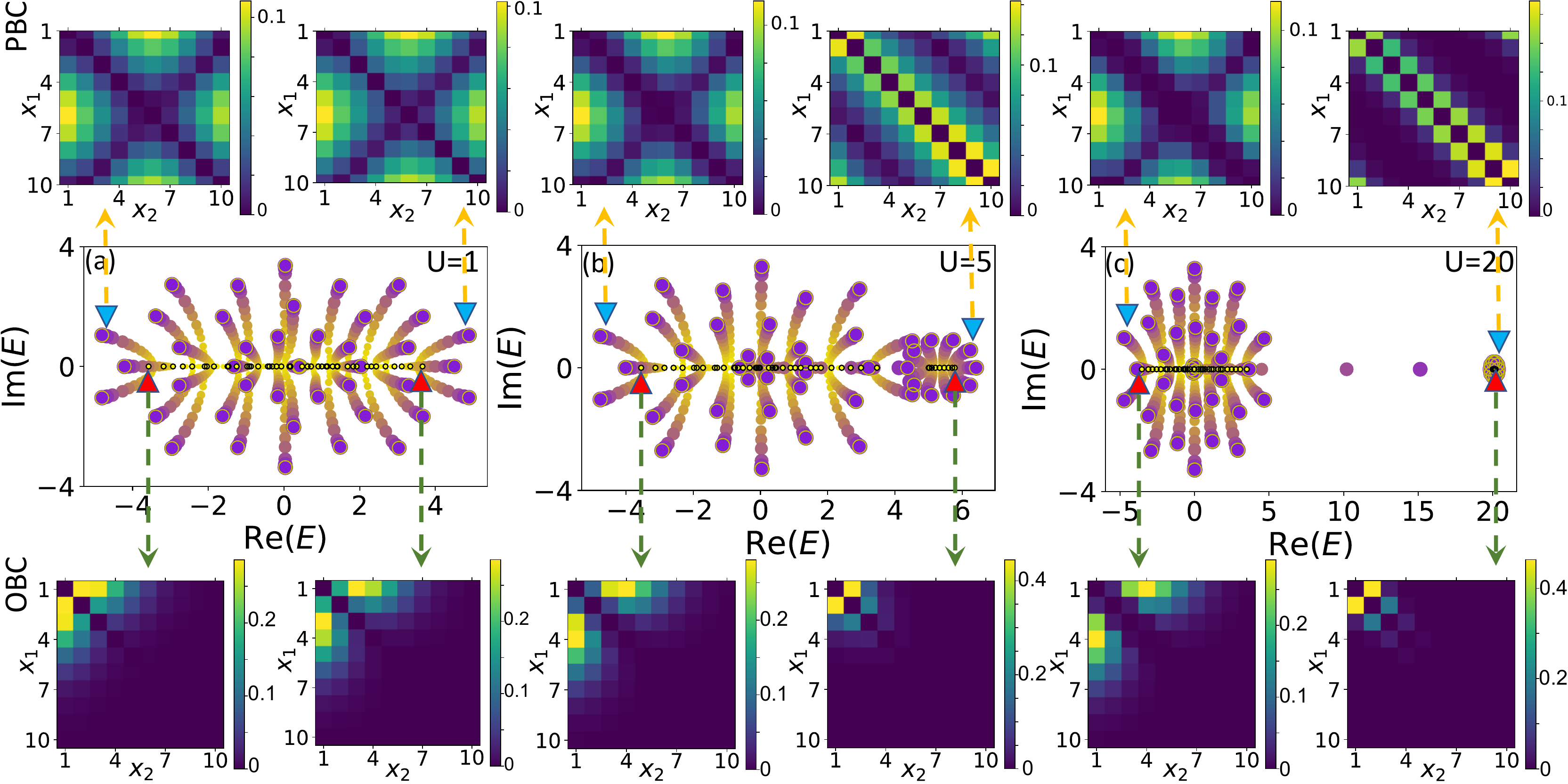}
\caption{{(a)/(b)/(c)  PBC vs. OBC spectra for 2 fermions with various NN repulsion $U=1/5/20$, and how the PBC-OBC spectral evolution reflects interplay between NN repulsion and asymmetric gain/loss pumping. In the center row, the PBC/OBC eigenenergies are depicted by circled purple/small yellow dots. PBC and OBC spatial profiles of the lowest/highest $Re(E)$ eigenstates, where $U$ is effectively attractive/repulsive, is shown on the top and bottom rows. Shorter PBC-OBC evolution curves correspond to more similar spatial localization lengths. We used $\alpha=1$ and $L=10$ throughout.}}
\label{fig_sf}
\end{figure*}

\noindent{\it Spectral analysis of interplay.--} Ultimately, the real space FS is a consequence of fermion accumulation {under OBCs, but not periodic BCs (PBCs)}.  To understand the {OBC} accumulation more deeply at the level of the entire spectrum, we interpolate between OBC and PBC by adiabatically turning on the hoppings $Je^{\pm \alpha}$ and repulsion $U$ between the first and the last sites from $0$ to $100\%$. In general, {PBC and OBC spectra and their respective eigenstate profiles can be drastically different}.    A longer OBC-PBC spectral flow trajectory implies stronger spatial accumulation~\cite{suppmat}, as shown in Fig.~\ref{fig_sf} for various NN repulsion strengths $U$ between 2 fermions, where yellow-purple curves connecting the PBC eigenenergies (circled purple dots) {collapse} onto real OBC eigenenergies (small yellow dots). In the weakly repulsive $U=1$ case, both PBC and OBC spectra are gapless (forming a single cluster), but larger $U$ repulsions leads to the formation of a high energy eigenenergy cluster (band), opening up a real OBC Mott gap.

The dispersion of this high energy band, as well as the insulating gap width, can be partially understood from the OBC-PBC spectral flow. In the large $U=20$ limit, {a high energy state experience strong effective attraction instead of repulsion, and} contains both fermions in adjacent sites. Since {localization property of} this configuration is largely agnostic to boundary conditions, we expect its PBC to OBC trajectories to be very short, with almost identical PBC and OBC eigenenergies and minute intraband dispersion. As $U$ decreases, the PBC-OBC flow trajectories from the higher band necessary get longer, and will eventually intersect with those from the lower band, as for $U=5$. This gives a scenario where the OBC spectrum is gapped while the PBC spectrum is critical, which we also see is inevitable from the spectral flow analysis.
\vspace{0.15cm}

\noindent{\it Proposed demonstration with cold atoms.--} 
Our emergent FS and competing CDW can be qualitatively realized in any fermionic lattice with effectively asymmetric hopping. Recently, it was realized that such nonreciprocal hopping can be effectively implemented (in a rotated representation) by introducing atom loss only~\cite{WangPRL2019,hofmann2019reciprocal,li2019coldatom}. This implementation is especially feasible in cold-atom systems trapped in optical lattices \cite{rev_cold_1,rev_cold_2}.

As an example, we propose a 2-level setup $H=H_0+H_{\rm nH}+H_{\rm int}$, with the Hermitian non-interacting part
\begin{eqnarray}
H_0&=&-\sum_j \left[it_q\hat{c}^\dagger_{j,\uparrow}\hat{c}_{j,\downarrow}+t_d(\hat{c}^\dagger_{i,\uparrow}\hat{c}_{j+1,\downarrow}-\hat{c}^\dagger_{i,\uparrow}\hat{c}_{j+1,\downarrow})\right.\nonumber\\
&&\left.+t_p(\hat{c}^\dagger_{j,\uparrow}\hat{c}_{j+1,\uparrow}-\hat{c}^\dagger_{j,\downarrow}\hat{c}_{j+1,\downarrow})\right]+h.c.,
\label{exp}
\end{eqnarray}
on-site dissipation $H_{\rm nH}=-ig\hat{c}^\dagger_{i,\downarrow}\hat{c}_{i,\downarrow}$ added to the pseudospin-down component, and $H_{\rm int}=\sum_j\sum_{\sigma,\sigma'}Un_{j,\sigma}n_{j+1,\sigma'}$ ($\sigma=\ \uparrow,\downarrow$) being the NN interaction tunable by Feshbach resonance \cite{rev_cold_1,rev_cold_2}.
Note that $H_0$ has already been realized in a resonantly driven 1D optical lattice with ''two-tone" periodic lattice shaking, where the pseudospin is simulated by {two orbitals}~\cite{jin2019Creutz}. $H_{\rm nH}$ can be induced by resonant exciting of the atoms out of their orbitals~\cite{Jiaming2019gainloss}. {The atomic density $\rho(x)$ can be experimentally resolved at atomic resolution \cite{exp1,exp2,exp3}, and therefore experiments can indeed probe the interplay between the CDW and the FS localization.   Clearly then, our proposal is already within reach of today's experimental techniques}. 


\begin{acknowledgements}
\emph{Acknowledgements.}-- J. G. acknowledges support from the Singapore NRF grant No.~NRF-NRFI2017-04 (WBS No.~R-144-000-378-281). Many-body Hamiltonians were constructed using QuSpin \cite{quspin1,quspin2}. We thank Xizheng Zhang, Da-Jian Zhang, Longwen Zhou, {Ronny Thomale, Titus Neupert and Emil Bergholtz} for discussions.

\end{acknowledgements}


\clearpage
\pagebreak
\newpage

\onecolumngrid
\begin{center}
\textbf{\large Supplemental Online Material for ''Emergent Fermi surface in a many-body non-Hermitian Fermionic chain" }
\end{center}
\date{\today}

\vspace {0.1cm}
{\small This supplementary contains the following material arranged by sections:\\
\begin{enumerate}
\item Analytic derivations and numerical results for the spatial density distribution, which exhibits a ''Fermi surface'' discontinuity for sufficiently large hopping asymmetry.
\item Entanglement entropy results on this emergent Fermi surface, which interrogates its unconventional properties via finite size, finite temperature and hopping asymmetry scalings. 
\item PBC-OBC spectral evolution in terms of the evolution of the many-body eigenstate profile.

\end{enumerate}
}

\setcounter{equation}{0}
\setcounter{figure}{0}
\setcounter{table}{0}
\setcounter{page}{1}
\setcounter{section}{0}
\setcounter{secnumdepth}{3}
\renewcommand{\theequation}{S\arabic{equation}}
\renewcommand{\thefigure}{S\arabic{figure}}
\renewcommand{\thesection}{S\Roman{section}}
\renewcommand{\thepage}{S\arabic{page}}
\vspace {1cm}

\section{Spatial density distribution and real space Fermi surface}

\subsection{Analytic derivation for non-interacting case}
\subsubsection{Single-body exact solution for OBCs}
\noindent Consider our Hamiltonian (Eq.~(\ref{pm}) of the main text) in the non-interacting limit:
\begin{equation}
H=J\sum_{i} a\,c^\dagger_{i+1}c_i + a^{-1}c^\dagger_i c_{i+1},
\end{equation}
where $a=e^{\alpha}$. Under PBCs with a system length of $L$ sites, the solution is obvious upon writing the Hamiltonian in momentum space : $H(k)=ae^{ik}+a^{-1}e^{-ik}$, with $k=2\pi j/L$, $j\in {1,...,L}$. Under OBCs, the (right) eigensolutions for this simple Hamiltonian $H$ still possesses an exact analytic form: $H\psi_j(x)=\varepsilon_j\psi_j(x)$, where
\begin{subequations}
\begin{equation}
\varepsilon_j= 2J\cos\frac{\pi j}{L+1},
\label{energy}
\end{equation}
\begin{equation}
\psi_j(x) = \mathcal{N}_ja^{-x}\sin\frac{\pi j x}{L+1},
\end{equation}
\end{subequations}
with the normalization constant given by
\begin{equation}
\mathcal{N}_n=\sqrt{\frac{(a^2-1)\left(4a^2+(a^2-1)^2\text{csc}^2\frac{n\pi}{L+1}\right)}{a^2(a^2+1)\left(1-a^{-2(L+1)}\right)}}.
\label{norm}
\end{equation}
It is not hard to show that in the Hermitian limit of $a\rightarrow 1$, $\mathcal{N}_j\rightarrow \sqrt{\frac{2}{L+1}}$. Note that the $\mathcal{N}_j$ above corresponds to the normalization convention defined by $\sum_x|\psi_j(x)|^2=1$; if we had chosen the biorthogonal norm instead, the $a^{-x}$ factor of the right eigenvector will cancel with the $a^{x}$ factor of the left eigenvector, resulting in a more trivial normalization factor of $\sqrt{\frac{2}{L+1}}$. The biorthogonal norm, however, is the overlap between $H$ and $H^\dagger$ right eigenstates, which is impervious to skin effect physics. Therefore, we shall only consider the normalization between the right eigenstates, for which Eq.~(\ref{norm}) can be proven by expressing the normalization sum as a geometric series involving powers of $a$ and $e^{\pm \frac{i\pi j }{L+1}}$.

\subsubsection{Many-body groundstate}
\begin{figure}
\centering
\subfloat[]{\includegraphics[width=.24\linewidth]{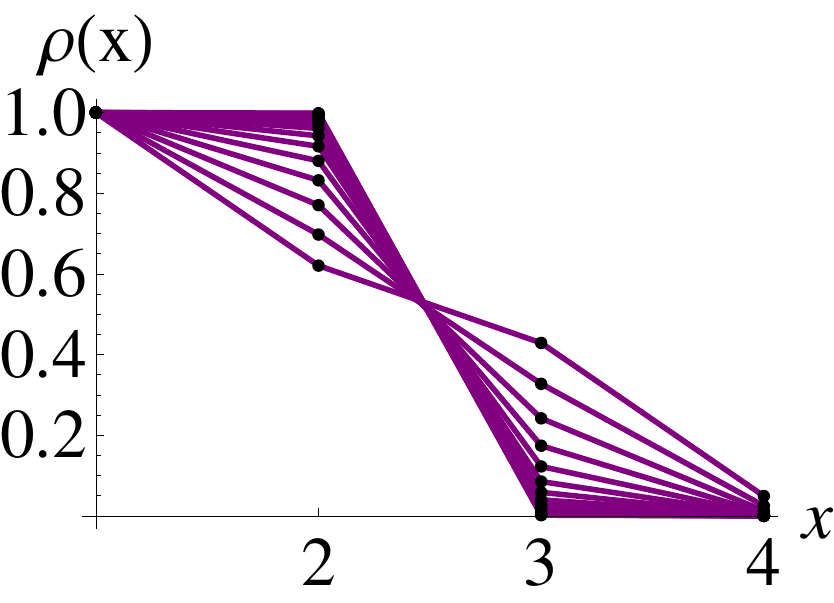}}
\subfloat[]{\includegraphics[width=.24\linewidth]{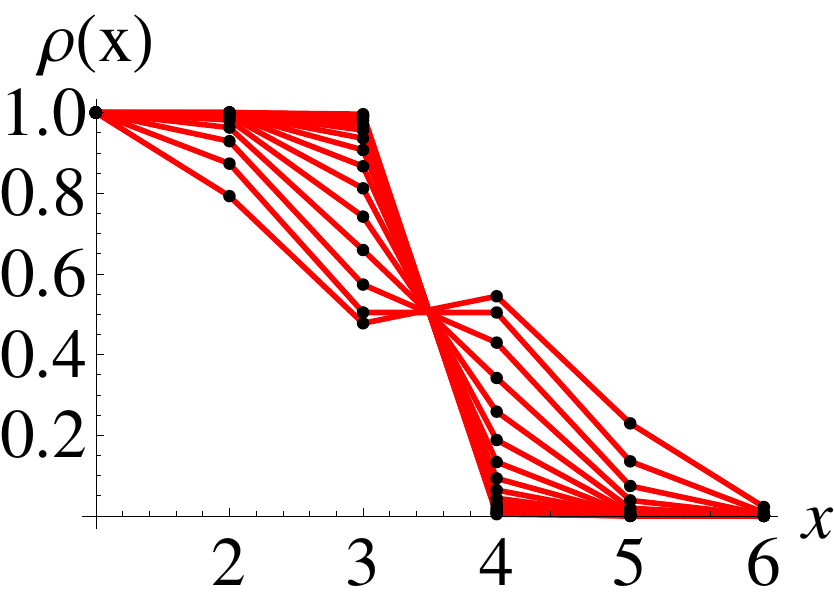}}
\subfloat[]{\includegraphics[width=.24\linewidth]{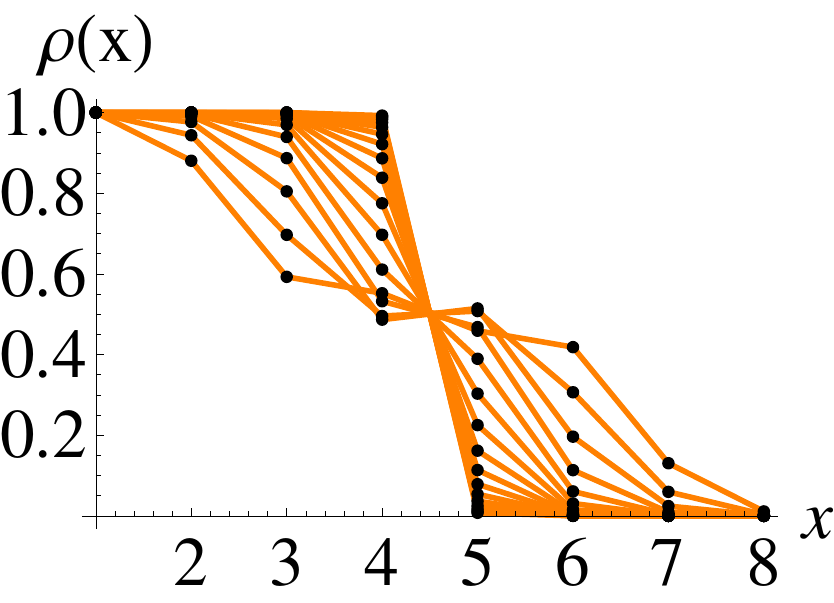}}
\subfloat[]{\includegraphics[width=.25\linewidth]{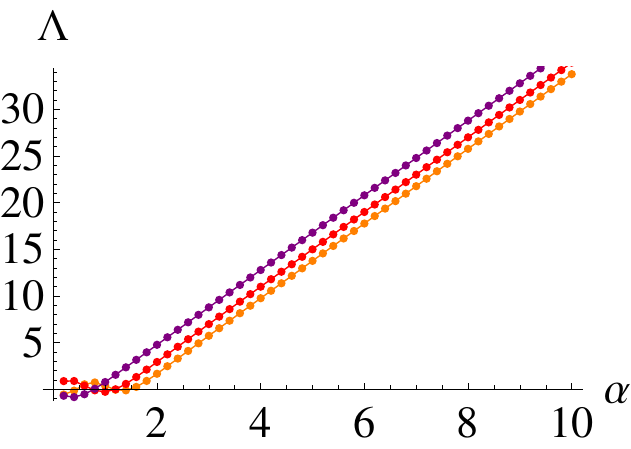}}
\caption{(a-c) Spatial density profiles $\rho(x)$ for $n=2,3,4$ particles with hopping asymmetries ranging from $\alpha=1$ to $4$ at intervals of $0.2$. The jumps become sharper with increasing $\alpha$. (d) The linear relation between the Fermi-Dirac distribution parameter $\Lambda$ Eq.~(\ref{FD}) and hopping asymmetry $\alpha$, at least for $\alpha>2$. Purple/red/orange curves represent $n=2,3$ and $4$ cases at half filling, as depicted in (a-c).  }
\label{fig:skin}
\end{figure}

From Eq.~(\ref{energy}), we know that $\varepsilon_j$ decreases monotonically with $n$, so that the $N$-particle fermionic ground state is the Slater determinant of $\psi_L$, $\psi_{L-1}$,...,$\psi_{L-N+1}$ (Excited many-body eigenstates can be obtained by choosing $\psi_j$'s that do not correspond to the lowest allowed energies.). Specifically, we have for the ground state
\begin{eqnarray}
\psi(x_1,x_2,...,x_N)&=&\frac1{\sqrt{N!}}\sum_\sigma \text{sgn}(\sigma)\prod_{j=1}^N \psi_{L-j+1}(x_{\sigma(j)})\notag\\
&=&\frac{\Omega_N}{\sqrt{N!}}a^{-\sum_{j=1}^Nx_j}\left[\sum_\sigma \text{sgn}(\sigma)\prod_{j=1}^N(-1)^{x_{\sigma(j)}+1}\sin\frac{\pi j x_{\sigma(j)}}{L+1}\right]\notag\\
&=&\frac{(-1)^N\Omega_N}{\sqrt{N!}}(-a)^{-\sum_{j=1}^Nx_j}\left[\sum_\sigma \text{sgn}(\sigma)\prod_{j=1}^N\sin\frac{\pi j x_{\sigma(j)}}{L+1}\right]\notag\\
&=&\frac{(-1)^N\Omega_N}{\sqrt{N!}}(-a)^{-Nx_{\rm CM}}\tilde \psi(x_1,x_2,...,x_N)
\label{psi1}
\end{eqnarray}
where $\Omega_N=\prod_{j=1}^N\mathcal{N}_{L-j+1}$ is the overall normalization constant. We see that $\psi(x_1,x_2,...,x_N)$ factorizes into the product of the Hermitian ($a=1$) Slater determinant $\tilde \psi(x_1,x_2,...,x_N)$ and an exponentially decaying envelop controlled by center-of-mass coordinate $x_{\rm CM}=\frac1{N}\sum_{j=1}^Nx_j$. 
Although the hopping asymmetry $a$ seems to enter only trivially in exponential factor, it in fact affects the ground state single-particle marginal probability density
\begin{equation}
\rho(x)\propto \int |\psi(x_1,...,x_{N-1},x)|^2 \prod_{j=1}^{N-1}dx_j
\label{rho1}
\end{equation}
nontrivially by creating a jump in the spatial profile reminiscent of a Fermi-Dirac distribution. This spatial density can always be exactly computed via Eqs.~(\ref{psi1}) and (\ref{rho1}), and is explicitly plotted for $n=2,3,4$ particles in Figs.~\ref{fig:skin}(a-c) for $L=2n$ and $\alpha=\log a =1,1.2,1.4,...,4$. Since the $\psi_j$'s are in general not orthogonal in this non-Hermitian setting, the Slater determinant is not automatically normalized. As such, $\rho(x)$ as plotted had been normalized such that it sums to $n$, the number of fermions.

From Fig.~\ref{fig:skin}(a-c), nonzero $\alpha$ is seen to push the fermions to the left, resulting in an accumulation of charge. Due to the Pauli exclusion principle enforced by the Slater determinant, $\rho(x)$ is bounded above unity, thereby resulting in a jump in density between $x=n$ to $x=n+1$. This jump can be very well fitted by the Fermi-Dirac distribution
\begin{equation}
\rho_{\rm FD}(x)=\frac1{1+e^{\Lambda(x-n-1/2)}}
\label{FD}
\end{equation}
from which $\tanh\frac{\Lambda}{4}$ gives the gradient of the jump. In Fig.~\ref{fig:skin}(d), $\Lambda$ is numerically verified to be linearly related to $\alpha$ i.e. $\Lambda= 4\alpha+\Lambda_0(n)$, where $\Lambda_0(n)$ is a $n$-dependent offset. This linear relation however starts to break down at $\alpha<2$, where the asymmetric hoppings are too weak to ensure a monotonic spatial density distribution.

To analytically probe the origin and breakdown of the linear dependence of $\Lambda$ with $\alpha$, we consider the exact expression for $\rho(x)$ in the simplest possible analytically tractable case with $n=2$ and $L=4$. Eq.~(\ref{rho1}) gives
\begin{equation}
\rho(1)=\frac{a^6+4a^4}{1+4a^2+4a^4+a^6};\quad  \rho(2)=\frac{a^6-a^4+5a^2}{1+4a^2+4a^4+a^6};\quad \rho(3)=\frac{1-a^2+5a^4}{1+4a^2+4a^4+a^6};\quad \rho(4)=\frac{1+4a^2}{1+4a^2+4a^4+a^6},
\label{FDexact}
\end{equation}
for which the jump is given by $\rho(2)-\rho(3) = \frac{(a^2-1)(1-5a^2+a^4)}{1+4a^2+4a^4+a^6}=1-10/a^2+42/a^4 + \mathcal{O}(a^{-6})$. Comparing this asymptotic result with that of the Fermi-Dirac jump $\tanh\frac{\Lambda}{4}\approx 1-2e^{-\Lambda/2}$ in the large $\Lambda$ limit, we see that $\Lambda$ must thus scale like $4\log a = 4\,\alpha$.

From the expressions of $\rho(x)$ given by Eq.~(\ref{FDexact}), we also see that a kink develops when $a=e^\alpha< \sqrt{\frac{5+\sqrt{21}}{2}}=2.1889$, such that $\rho(3)>\rho(2)$. This kink extrapolates down to the Hermitian limit of $a=1$, where the oscillatory nature of the ground state triumphs any asymmetric pumping. Although this analysis was performed for the simplest case, its results hold true qualitatively for generically many fermions.

Although we have identified the density profile with a Fermi-Dirac distribution, $\Lambda$ is still not a true inverse temperature parameter because it is conjugate to a position variable and not energy. But just like how ordinary Fermi seas arise from the balance between universal energy minimization and QDP, an emergent \emph{real space} Fermi sea is formed when asymmetric gain/loss relentlessly pushes all fermions in one direction until counteracted by QDP. 
Indeed, the real space FS is exactly analogous to a zero temperature FS in the $|\alpha|\rightarrow \infty$ limit, where all fermions completely condense onto one side. But for finitely large $\alpha$ where $\rho(x)$ is not a perfect step function, the identification with a finite temperature FS is more subtle. After all, the spatial eigenstate profile is a pure-state property, but temperature is a characteristic of a mixed-state ensemble. A more careful identification of the temperature of this emergent Fermi surface will be given in the following section by invoking entanglement entropy.

\subsection{Numerical results for general cases}

More generally, we obtain the spatial density in the presence of interactions and nonzero temperature through exact diagonalization. For the latter, we assume that the system is in an incoherent ensemble of the eigenstates, such that $\rho(x)=\sum_j e^{-\beta E_j}\rho_j(x)$, where $E_j$ and $\rho_j(x)$ are respectively the energy and spatial density of the $j$-th many-body right eigenstate.

\begin{figure}
\includegraphics[width=.94\linewidth]{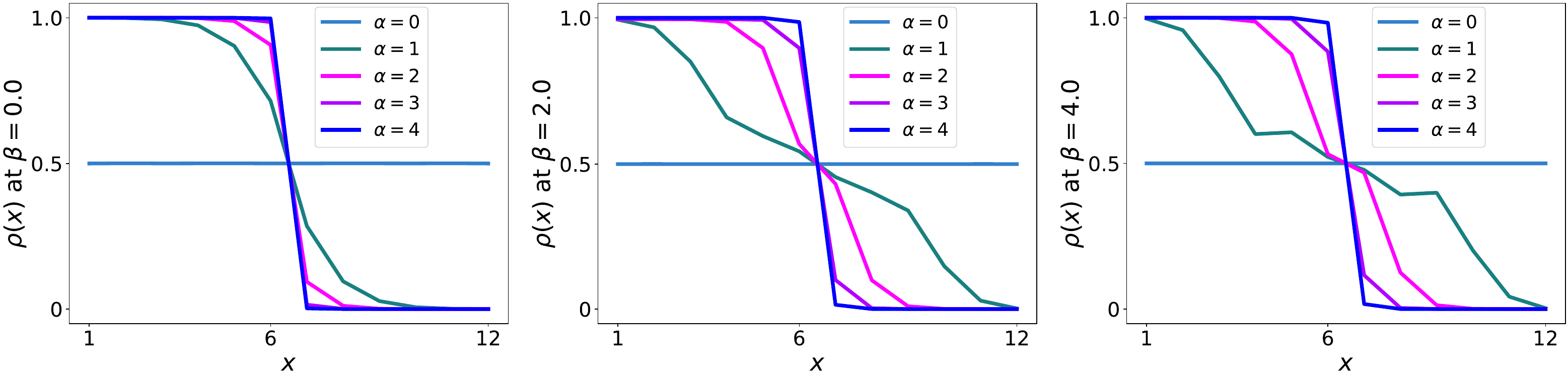}
\caption{Spatial densities at different inverse temperatures $\beta$, all at half filling $L=2n=12$ and in the noninteracting limit $U=0$. As temperature decreases from left to right, $\rho(x)$ becomes less smooth due to the averaging over fewer accessible states, until it ultimately reduces to the ground state distribution which is well approximated by $\beta=4.0$. }
\label{fig_s1}
\end{figure}

The effects of finite temperature are shown in Fig.~\ref{fig_s1}, still without interactions for easy comparison with the zero temperature (ground state) results of Fig.~\ref{fig:skin}. Evidently, at small gain/loss asymmetry $\alpha$, the effect of asymmetric pumping increases with temperature (smaller $\beta$), since more states become thermally accessible. At larger $\alpha>2$, the temperature no longer affects $\rho(x)$ so much since the Fermi surface is already well-defined, consistent with the entanglement entropy results in Fig.~\ref{fig_ee}(b) of the main text where the EE remains almost independent of $\beta$. In general, higher temperatures lead to a smoother $\rho(x)$ by involving more states. However, by comparing Figs.~\ref{fig:skin} and \ref{fig_s1}, we see that the ground state profile is already well approximated by a temperature of $\beta^{-1}=1/4$, which is actually already very low in comparison with the large energies associated with the $Je^{\alpha}$ hoppings.

\begin{figure}
\includegraphics[width=.94\linewidth]{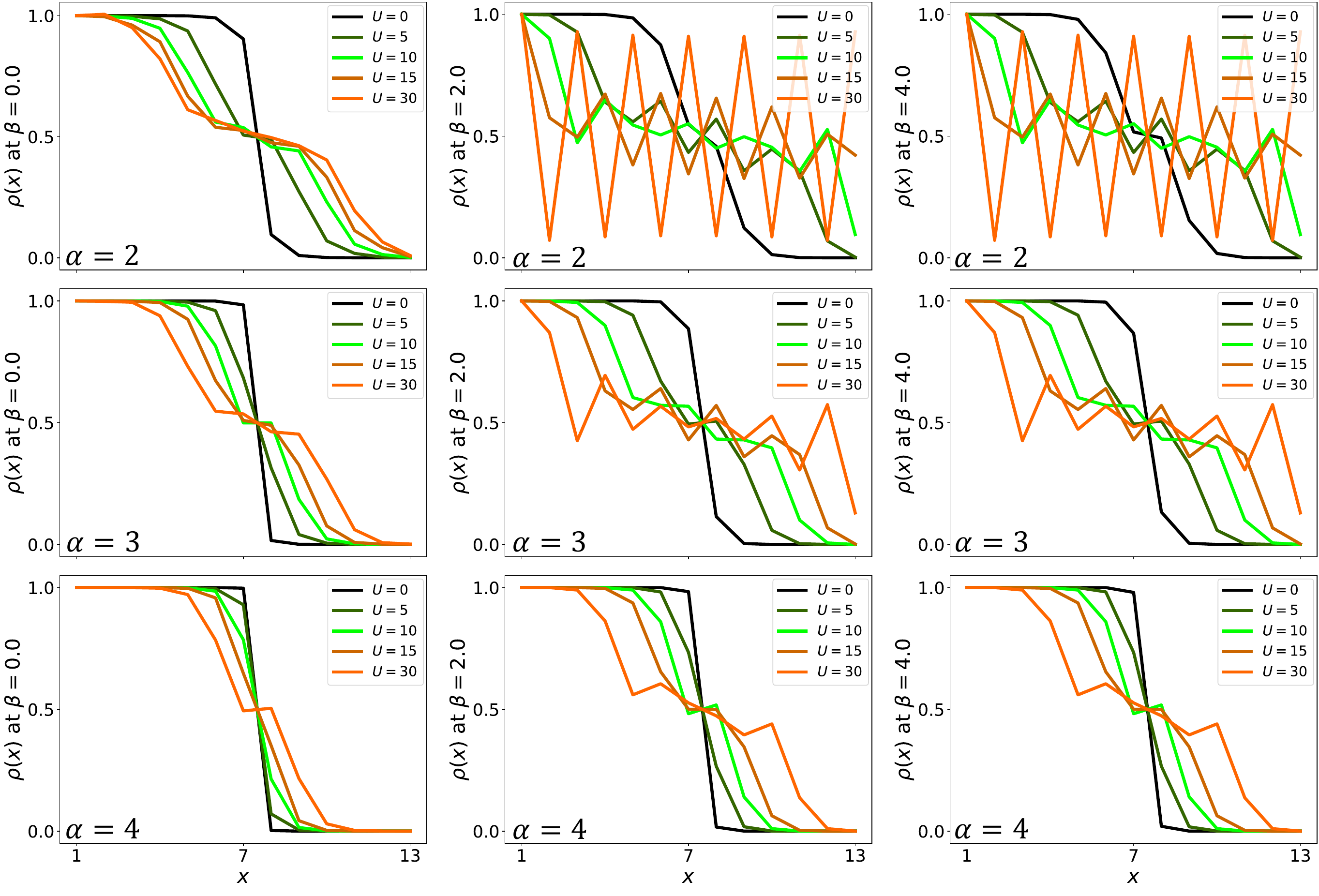}
\caption{Spatial densities at different inverse temperatures $\beta$ and NN repulsive interaction strengths $U$, all at approximately half filling $n=7,\, L=13$. The interactions favor CDWs, which are more pronounced at lower hopping asymmetries $\alpha$. However, higher temperatures (smaller $\beta$) make higher energy states that penalized by the interactions also accessible, thereby drowning out the effects of CDWs in $\rho(x)$. }
\label{fig_s1b}
\end{figure}

Next, we present the interplay of finite temperature with interaction effects ($U>0$) in Fig.~\ref{fig_s1b}, complementing Fig.~\ref{fig_u}(a) of the main text. At finite temperatures (middle and right column), we observe spatially oscillating charge density waves (CDWs) that exist to offset the energy penalty associated with nearest neighbor repulsion. These CDWs, which are dominant in the Hermitian limit, are suppressed when the asymmetric pumping is sufficient large ($\alpha>3$). As temperatures increase, higher energy states that are not penalized by the NN repulsion also become accessible, thereby leading to decreased contributions by the CDWs. In the infinite temperature limit (left column), CDWs are totally overwhelmed by the higher energy states, such that the interactions only smudge the Fermi surface in their competition with asymmetric pumping effects.

\section{Entanglement entropy computations}

\subsection{Generalities}
Without considering interactions $U$, our system consists of free fermions, and the entanglement spectrum and entropy can be extracted from the correlation matrix. From \cite{Peschel2009}, Wick theorem states that every fermionic correlation function can be expressed as a linear combination of products of two-point correlators, i.e. $C_{ij}=\langle \psi |c^{\dagger}_ic_j|\psi\rangle$, where $c^{\dagger}_i,c_j$ are restricted to a subsystem A defined by the entanglement cut. Therefore, the reduced density matrix of subsystem $A$ must take the Gaussian form $\rho_A = \frac{1}{Z_A} e^{-h_A}$, where $h_A=\sum c_l^{\dagger}h_{lk}c_k$ is referred as the entanglement Hamiltonian and $Z_A$ is the reduced partition function of A. One can simultaneously diagonalize $C$ and $h_A$ and find $C=\frac{e^{-h_A}}{1+e^{-h_A}}$. The entanglement entropy is thus given by
\begin{eqnarray}
S_{\rm ent} &=& -{\rm Tr}[\rho_A \log \rho_A] = -{\rm Tr}[\frac{e^{-h_A}}{1+e^{-h_A}}\log \frac{e^{-h_A}}{1+e^{-h_A}} + \frac{1}{1+e^{-h_A}}\log \frac{1}{1+e^{-h_A}}]\\
		  &=& -\sum_m[\lambda_m\log \lambda_m+(1-\lambda_m)\log(1-\lambda_m)]
\end{eqnarray}
where $\lambda_m$ are the eigenvalues of $C$. For a non-Hermitian system, this approach works for either LL/RR or LR correlators viz. $C_{ij}=\langle \psi^L|c^{\dagger}_ic_j|\psi^L\rangle$, $C_{ij}=\langle \psi^R|c^{\dagger}_ic_j|\psi^R\rangle$ or $C_{ij}=\langle \psi^L|c^{\dagger}_ic_j|\psi^R\rangle$. But since we are only interested in the entanglement properties of the right eigenstate, we shall only employ the RR choice in this work.

For the more generic interacting cases we studied, the EE has to be computed by working in the full Hilbert space of the system. For illustration, consider a fermionic chain consisting of $L$ sites. We can represent it as a chain of $L$ qubits, with $|0_i\rangle$ and $|1_i\rangle$ corresponding to an empty/occupied site $i$. Note that if we only consider particle-conserving systems, the physical Hilbert space is actually a subspace of this full Hilbert space. We partition this chain of qubits into left and right parts, and consider the left part as subsystem $A$. After solving the non-Hermitian Hamiltonian defined on this chain, we can construct a well-defined density matrix of the system from the right eigenstates via $[\rho^{RR}]_{\mu\nu}=|\psi^R_\mu\rangle\langle \psi^R_\nu|$. Next we trace out degrees of freedom of the complement of a subsystem $A$ and obtain a reduced density matrix $\rho_A^{RR}=\mathrm{Tr}_{A^{\rm c}}|\psi^R\rangle\langle\psi^R|=\sum_r \lambda_r^2 |\psi^R_{r,A}\rangle\langle \psi^R_{r,A}|$, where $|\psi^R\rangle=\sum_r\lambda_r |\psi^R_{r,A}\rangle\otimes |\psi^R_{r,A^{\rm c}}\rangle$ is the Schmidt decomposition of a representative $|\psi^R\rangle$ and $A^{\rm c}$ is the complement of $A$. For some elementary intuition~\cite{matsueda2012holographic,lee2014exact}, the simplest example of a Schmidt decomposition can be found in the SVD decomposition of a matrix, where the original matrix is decomposed into a linear combination of rank one matrices, each which is a tensor product of a row/columns vectors belonging to $A$ and $A^{\rm c}$.

The von Neumann EE with respect to the entanglement cut separating $A$ and $A^{\rm c}$ is similarly given by
\begin{eqnarray}
S_{\rm ent} &=& -\mathrm{Tr}[\rho^{RR}_A\mathrm{log}\rho^{RR}_A]\notag\\
&=&-\sum_r [\lambda_r\log\lambda_r + (1-\lambda_r)\log(1-\lambda_r)].
\label{sent}
\end{eqnarray}
where $\lambda_r$ are the eigenvalues of $\rho_A$. Note that this latter approach is much more resource consuming than the correlation matrix approach, as diagonalization and partial trace are done on matrices of size $2^L$, where $L$ is the system size.

\subsection{Entanglement entropy results for non-interacting cases}
\subsubsection{Scaling of $S_{\rm ent}$ in the Hermitian non-interacting limit ($\alpha=U=0$)}

\begin{figure}[H]
\centering
\includegraphics[width=0.85\linewidth]{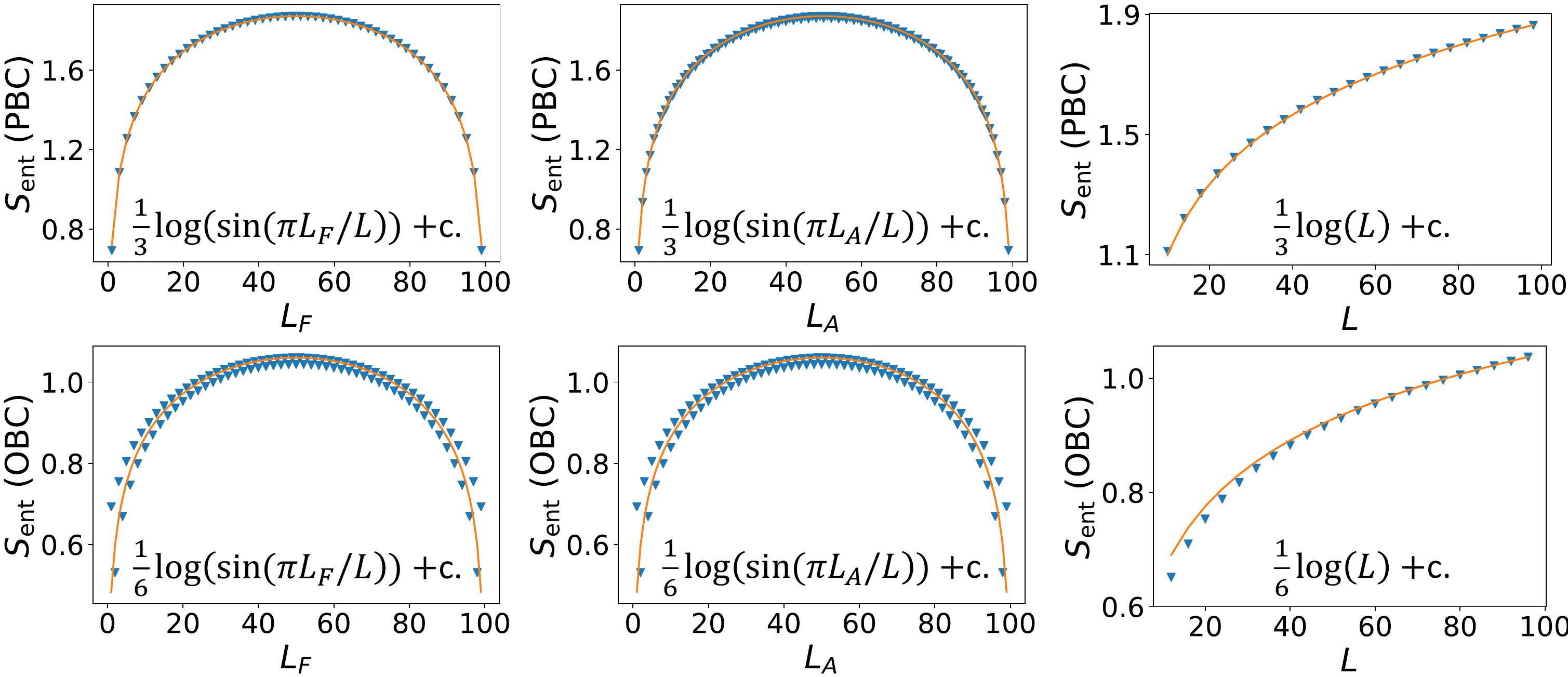}
\caption{Scaling behavior of the ground state EE with $L_F$, $L_A$ and $L$ for the $U=\alpha=0$ case, as computed via the correlation matrix. We have set $L=100$ and $L_A=L_F=L/2$ when they are not being varied. $L_A$ and $L_F$ are implemented as position and momentum space ''cuts'', i.e. by projecting onto the first $L_A$ sites and $L_F$ lowest eigenstates respectively. Orange curves denote the fits from Eq.~(\ref{PRPRP}).
}
\label{fig_s2a}
\end{figure}

\begin{figure}
\includegraphics[width=0.85\linewidth]{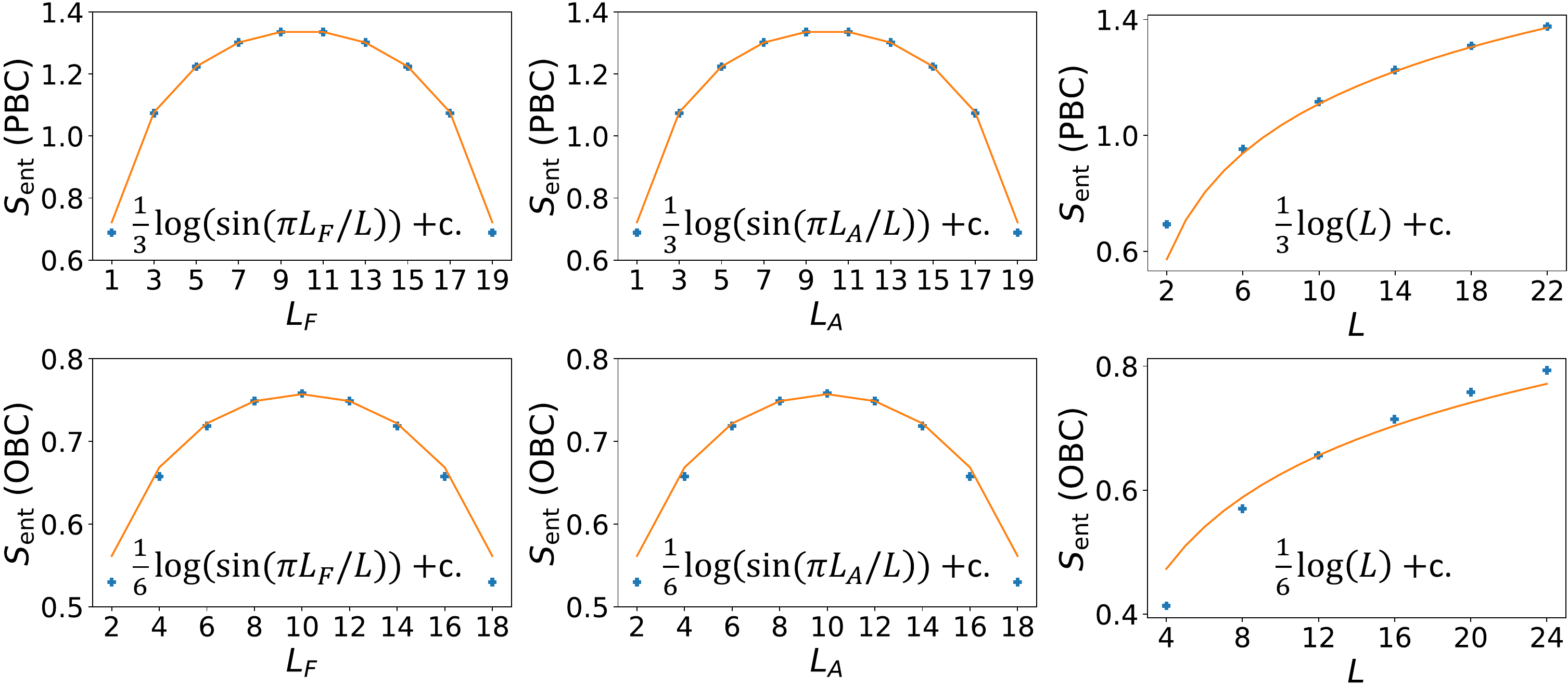}
\caption{Scaling behavior of the ground state EE with $L_F$, $L_A$ and $L$ for the $U=\alpha=0$ case, as computed by considering the full Hilbert space. We have set $L=20$ and $L_A=L_F=L/2$ when they are not being varied. Orange curves denote the fits from Eq.~(\ref{PRPRP}).}
\label{fig_s2b}
\end{figure}

\begin{figure}
\includegraphics[width=0.85\linewidth]{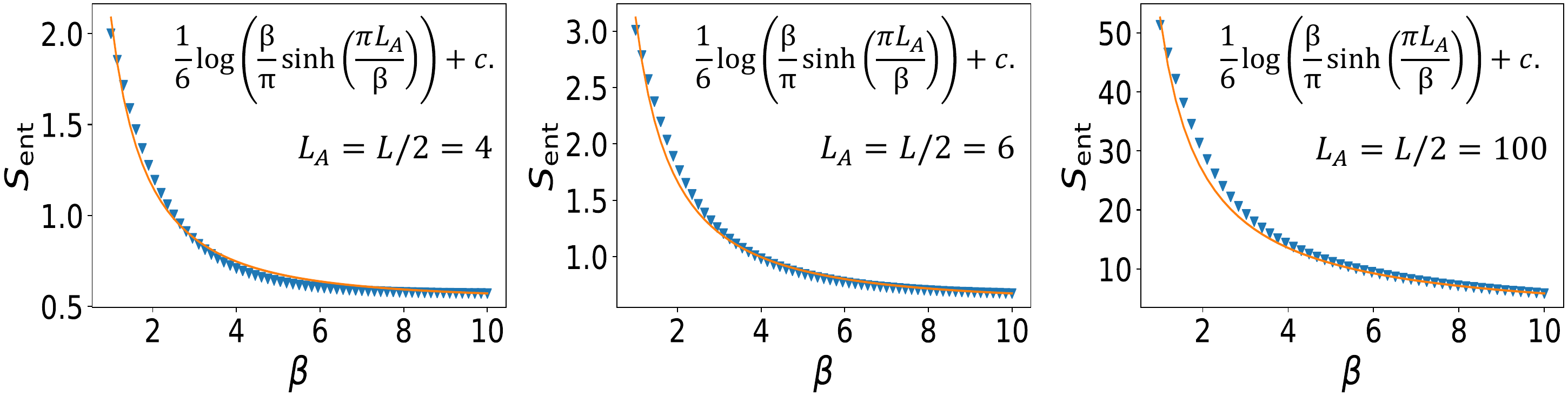}
\caption{Scaling behavior of OBC EE with the inverse temperature $\beta$ at half filling for the $U=\alpha=0$ case, as computed by considering the full Hilbert space for $L=8$ and $L=12$, and using the correlation matrix method for $L=200$. Orange curves denote the fits from Eq.~(\ref{obc_ee_t}) of the main text.}
\label{fig_s2}
\end{figure}

We first benchmark both our correlation matrix and full Hilbert space EE computation approaches against known results for the simplest non-interacting Hermitian case at zero temperature. For such a critical fermionic chain $H=J\sum_x (c^\dagger_{x+1}c_x+c^\dagger_xc_{x+1})$, it is known that the EE scales like \cite{Lee2014PRPRP}

\begin{align}
S_{\rm ent}(\text{PBC}) \sim \frac{c}{3}\log\left[\frac{L}{\pi}\sin\frac{\pi L_A}{L}\sin\frac{\pi L_F}{L}\right]+\mathrm{const'.},\notag\\
S_{\rm ent}(\text{OBC}) \sim \frac{c}{6}\log\left[\frac{L}{\pi}\sin\frac{\pi L_A}{L}\sin\frac{\pi L_F}{L}\right]+\mathrm{const'.},
\label{PRPRP}
\end{align}
where $L_A$ is the position of the entanglement cut and $L_F/L=n/L$ is the filling fraction. As in Ref.~\cite{alexandradinata2011trace,Lee2014PRPRP,lee2015free}, we have re-interpreted $n$ as a ''momentum space cut'' $L_F$ such as to exploit the position-momentum symmetry of the model. As demonstrated in Figs.~\ref{fig_s2a} and \ref{fig_s2b}, our numerical results via both approaches exhibit excellent agreement with Eq.~(\ref{PRPRP}). {Then we benchmark our computation for OBC EE at finite temperature with the thermal density matrix $\rho_{A, \rm thermal}=\frac{1}{Z_A}\sum_j e^{-\beta E_j}\rho_{A,j}$. Our numerical results agree well with the conformally transformed standard results \cite{Cardy2004,Cardy2009} Eq.~(\ref{obc_ee_t}) of the main text as shown in Fig.~\ref{fig_s2}.}

\subsubsection{$\alpha$-dependence of non-interacting $S_{\rm ent}$ at different temperatures and system sizes $L$ }

\begin{figure}
\includegraphics[width=0.8\linewidth]{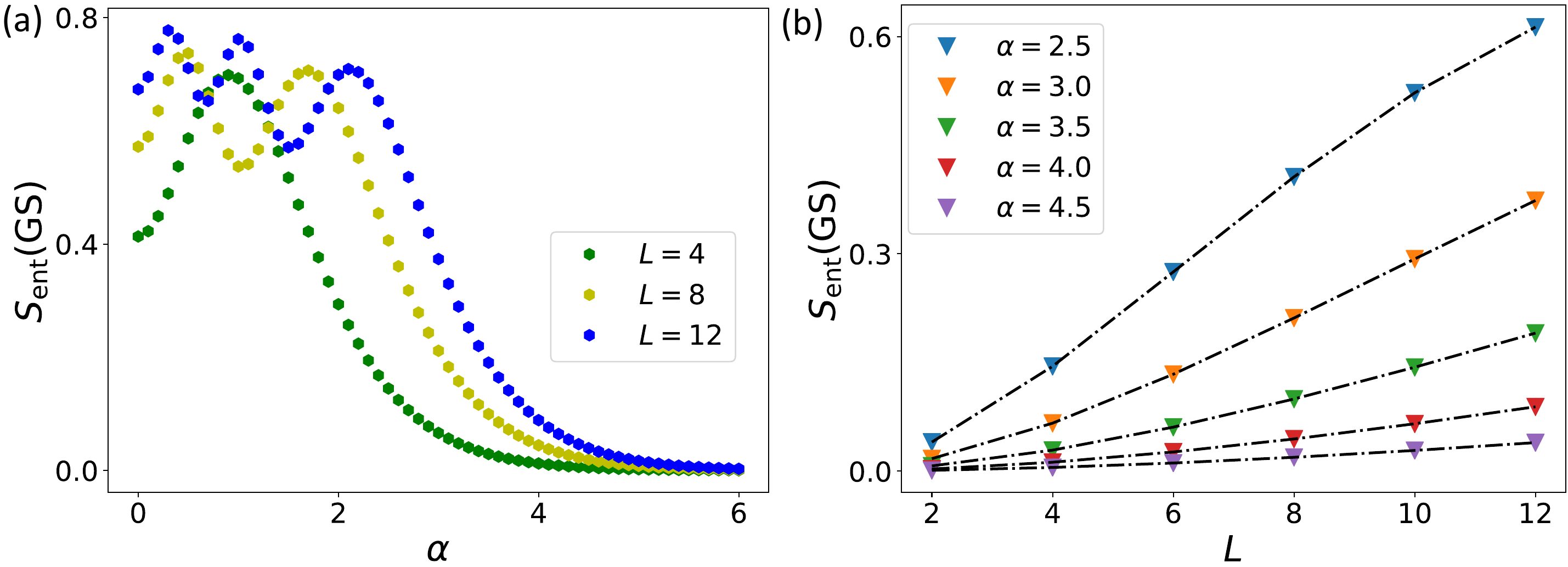}
\caption{(a) Ground state entanglement entropy $S_{\rm ent}(\text{GS})$ vs. $\alpha$ under OBCs with $U=0$. Universal behavior only exists for $\alpha>2$, where a well-defined real space  FS exists. The cuts are taken at $L_A=L/2$ at half-filling. (b) $S_{\rm ent}(\text{GS})$ vs. $L$ under OBCs with $U=0$, showing monotonic scaling behavior at nonzero $\alpha$ that is no longer logarithmic. }
\label{fig_s4a}
\end{figure}

We now proceed to study the effects of nonzero hopping asymmetry $\alpha$ on $S_{\rm ent}$, but still at $U=0$. Since the real space Fermi surface only exists when the fermions are confined by a boundary, we shall henceforth consider only OBCs. At small $\alpha$, the FS is not well-defined, as evident in the previous section, and the ground state EE scales rather irregularly with both $L$ and $\alpha$ (Fig.~\ref{fig_s4a}(a)). Larger $\alpha$, however, gives rise to sharper well-defined FS, leading to lower EE. Plotting some of the same data against $L$ in Fig.~\ref{fig_s4a}(b), we observe a clear monotonic increase of the ground state EE with $L$, although not necessarily logarithmically as in the Hermitian case (Figs.~\ref{fig_s2a} and \ref{fig_s2b}).

\begin{figure}
\includegraphics[width=0.7\linewidth]{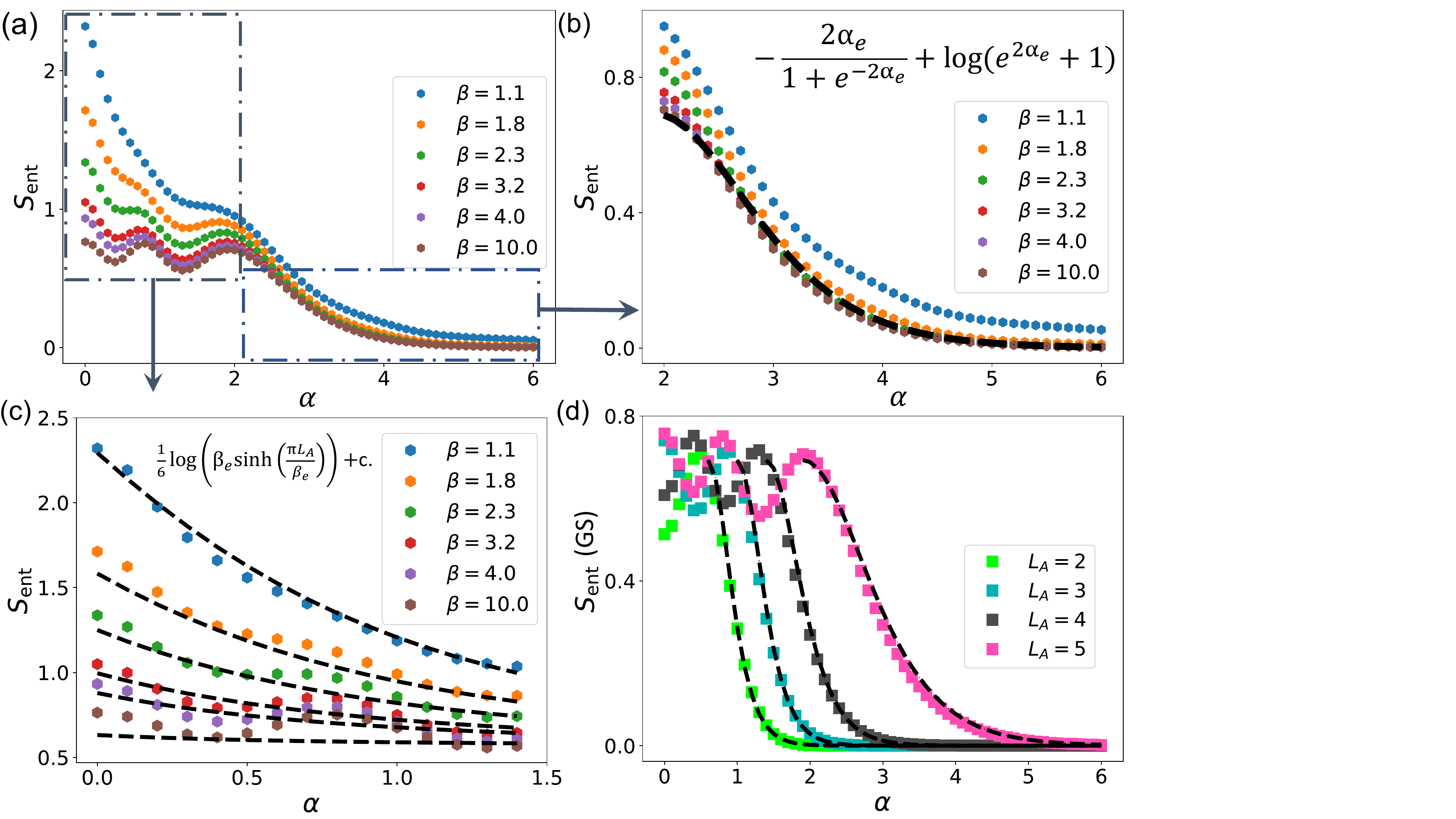}
\caption{(a) OBC Entanglement entropy vs. $\alpha$ at different inverse temperature $\beta$ with $L=2L_A=10$. (b) Fitting of data with $-2\alpha_{\rm e}/(1+e^{-2\alpha_{\rm e}})+\log(e^{2\alpha_{\rm e}}+1)$ where $\alpha_{\rm e}=\eta(\alpha-\alpha_0)$, $\alpha_0=1.9,\eta=1$.
(c) Fitting of data with $1/6\log(\beta_{e}/\pi\sinh(\pi L_A/\beta_{e}))$ where $\beta_e=(1+0.75\alpha)\beta+0.45\alpha^2$. (d) OBC Entanglement entropy vs. $\alpha$ at different subsystem length $L_A$ at $L=10$. Fitting of data with $-2\alpha_{\rm e}/(1+e^{-2\alpha_{\rm e}})+\log(e^{2\alpha_{\rm e}}+1)$ where $\alpha_0=\{0.6,1.0,1.4,1.9\}$ and $\eta=\{3.0, 2.7,2.0,1.0\}$ for $L_A=\{2,3,4,5\}$.}
\label{fig_s3a}
\end{figure}

\begin{figure}
\includegraphics[width=0.7\linewidth]{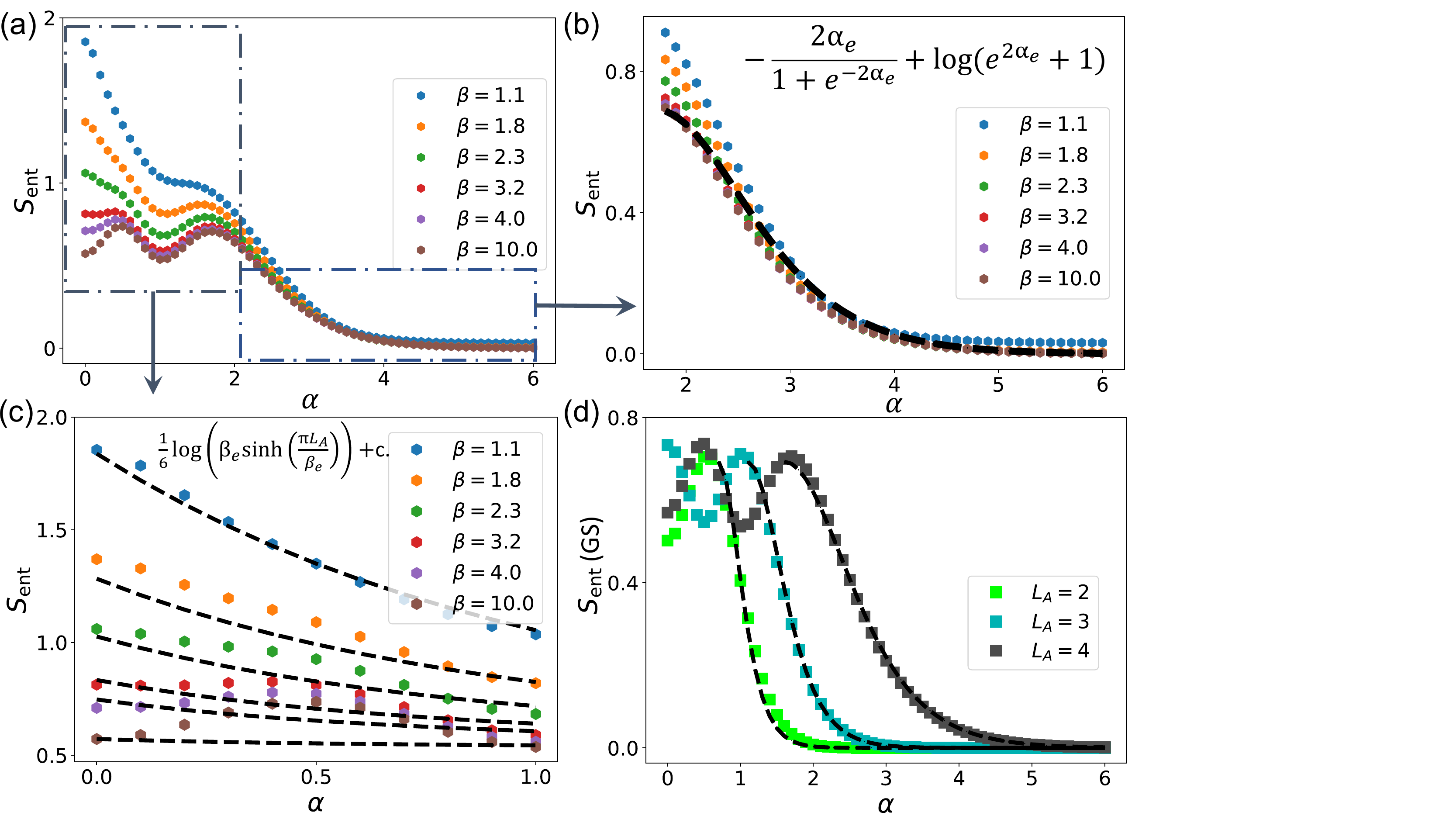}
\caption{(a) OBC Entanglement entropy vs. $\alpha$ at different inverse temperature $\beta$ with $L=2L_A=8$. (b) Fitting of data with $-2\alpha_{\rm e}/(1+e^{-2\alpha_{\rm e}})+\log(e^{2\alpha_{\rm e}}+1)$ where $\alpha_{\rm e}=\eta(\alpha-\alpha_0)$, $\alpha_0=1.6,\eta=1$. (c) Fitting of data with $1/6\log(\beta_{e}/\pi\sinh(\pi L_A/\beta_{e}))$ where $\beta_e=(1+0.75\alpha)\beta+0.25\alpha^2$. (d) OBC Entanglement entropy vs. $\alpha$ at different subsystem length $L_A$ at $L=8$. Fitting of data with $-2\alpha_{\rm e}/(1+e^{-2\alpha_{\rm e}})+\log(e^{2\alpha_{\rm e}}+1)$ where $\alpha_0=\{0.7,1.1,1.6\}$ and $\eta=\{3.0,1.9,1.0\}$ for $L_A=\{2,3,4\}$.}
\label{fig_s3b}
\end{figure}

To obtain further insight on the nature of the real space FS, we consider, as in the main text, nonzero temperatures which interplays with the smudging of the FS due to $\alpha$. Accompanying Fig.~2 of the main text which shows $S_{\rm ent}$ vs. $\alpha$ at various $\beta$ for $L=12$, here we present analogous results for $L=10$ and $L=8$ in Fig.~\ref{fig_s3a} and Fig.~\ref{fig_s3b} respectively. For large $\alpha>2$, the fitting to $\alpha_{\rm e}=\eta(\alpha-\alpha_0)$ of the main text holds very well too, consistent with our two-qubit caricature. For small $\alpha <2$, the fitting to Eq.~(2) of the main text also remains robust, with renormalized inverse temperature maintaining the form $\beta_e=(1+\epsilon\alpha)\beta+\delta\alpha^2$ with $\epsilon=\{0.75,0.75,0.75\}$ and $\delta=\{1.25,0.45,0.25\}$ for $L=\{12,10,8\}$.

\section{Further details on PBC-OBC spectral interpolation}

\begin{figure}
\includegraphics[width=1.0\linewidth]{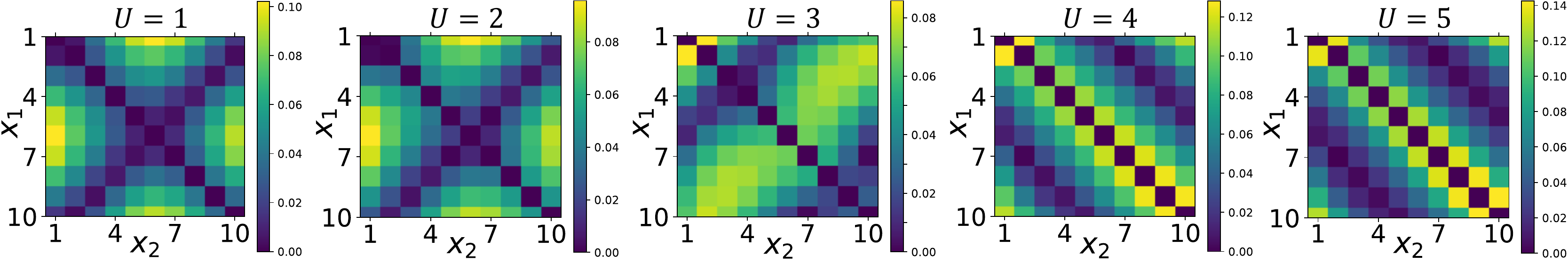}
\caption{Evolution of the PBC spatial density distribution of the two-fermion highest energy states as the interaction strength changes from $U=1$ to $U=5$, all with $L=10$ and $\alpha=1$.}
\label{fig_s8}
\end{figure}

Here we provide a more in-depth understanding of the insights gained from the PBC and OBC spectra of our model. In the PBC limit, translation invariance admits lattice momentum $q$ as a good quantum number, and allows our Hamiltonian to be recast into $H|_{PBC}=\sum_q J\cos (q+i\alpha)c^\dagger_qc_q+U\sum\limits_{\text{periodic }x}n_xn_{x+1} $. At the single-particle level, $\alpha$ thus appears as the imaginary part of the momentum that encodes, via $e^{i(q+i\alpha)x}\sim e^{-\alpha x}$, the non-Bloch spatial accumulation that will occur under OBCs. In other words, the extent of spatial accumulation $\alpha$ can be inferred from the difference between the PBC and OBC spectra, which are given, up to exponential corrections in system size~\cite{Lee2019anatomy}, by the energies corresponding to momenta $q+i\alpha$ and $q$ respectively. In a recent work relating OBC-PBC spectral flow to complex flux pumping~\cite{Lee2019anatomy}, the localization length can be computed from the imaginary gap band structure~\cite{he2001exponential,lee2016band}.

This still holds generally in many-body settings, with a longer spectral flow trajectory between the OBC and PBC spectra implying stronger spatial accumulation due to asymmetric gain/loss pumping, as presented in Fig.~4 of the main text. Note that in the weakly repulsive $U=1$ case, the PBC spectrum (circled purple dots) is almost identical to the non-interacting PBC spectrum $J(\cos (q+i\alpha) +\cos (q'+i\alpha))$, $q,q'\in [0,2\pi]$, which is symmetric about $Re(E)=0$. Negligible differences also exist between the lowest and highest energy OBC states (red triangles), since the physics is dominated by the asymmetric hoppings.

From the spatial profiles of the 2-fermion eigenstates in Fig.~4, we see that whether the two fermions are likely to be spatially close to each other depends a lot on the relative strength of the NN repulsion $U$, as well as the boundary conditions. In particular, the highest PBC energy state (where $U$ acts as an attraction instead of repulsion) depends drastically on $U$, as elaborated in Fig.~\ref{fig_s8}. As $U$ increases, the effective attraction between the fermions of the highest energy state encourages them to come close together, breaking the effective mirror symmetry of their joint spatial configuration at low $U$.


\bibliography{references_v7}

\end{document}